\documentclass[pra,twocolumn]{revtex4}%
\usepackage{amsfonts}
\usepackage{amsmath}
\usepackage{amssymb}
\usepackage{graphicx}%
\providecommand{\U}[1]{\protect\rule{.1in}{.1in}}

\begin{document}
\title{Entanglement of Gaussian states using beam splitter}
\date{\today}
\author{Rabia Tahira$^{1}$, Manzoor Ikram$^{1}$, Hyunchul Nha$^{2,*}$, and M. Suhail
Zubairy$^{1,2,3}$}
\affiliation{$^{1}$ Centre for Quantum Physics, COMSATS Institute of Information
Technology, Islamabad, Pakistan}
\affiliation{$^{2}$ Texas A\&M University at Qatar, Education City, P. O. Box 23874, Doha, Qatar}
\affiliation{$^{3}$ Institute of Quantum Studies and Department of Physics, Texas A\&M
University, College Station, Texas, 77843-4242, USA}

\begin{abstract}
We study an experimental scheme to generate Gaussian two-mode entangled states
via beam splitter. Specifically, we consider a nonclassical Gaussian state
(squeezed state) and a thermal state as two input modes, and evaluate the degree of entanglement at the output. 
Experimental conditions to generate entangled outputs are completely identified and the critical thermal noise to destroy entanglement 
is analytically obtained. 
By doing so, we discuss the possibility to link the resistance to noise in entanglement generation with the degree of single-mode nonclassicality.
\end{abstract}

\maketitle

\section{Introduction}

In quantum information processing more and more attention is directed to the
continuous variable (CV) systems as they have emerged as an alternative
resource to the discrete level systems. The CV states of considerable
importance are the Gaussian ones. The interest in this special class of states
stems from the experimental feasibility to produce them from reliable sources
\cite{Kimble} and to control them using accessible tools such as beam
splitters, phase shifter, and squeezers \cite{Reck}. 
The two-mode entangled
Gaussian states have been utilized in many of quantum
information applications \cite{Wang2007, Braunstein2005,Telep,Super,Key}.


Entanglement between two Gaussian modes is routinely generated in laboratory, e.g., the two output
beams of a nonlinear optical device (parametric down converter) \cite{Kimble}. 
Alternatively, a beam splitter, one of the linear optical devices, 
can also be used to generate quantum entanglement between two modes \cite{Book}. 
There have been many
studies for producing entanglement using beam splitter as an entangler
\cite{Tan, Sanders, Paris1999, Kim, Li, Marian2001}. In particular, Kim
\textit{et al.} studied the entangler properties with many different input
states, such as Fock states, pure- and mixed- Gaussian states. 
They conjectured that in order to obtain an entangled two-mode state
out of a beam splitter, it is necessary to have a nonclassical state at
one input, which was later proved in \cite{Wang}. Furthermore,
the \textit{sufficiency} of single-mode nonclassicality to generate entangled
states via beam splitter was demonstrated by Asboth \textit{et al.}
\cite{Asboth}. On another side, the separability criteria to detect such
entangled outputs via beam splitter have also been derived \cite{Zubairy,
nha}. Remarkably, a broad class of uncertainty inequalities was presented to
detect entanglement produced using generalized single-mode nonclassical
states, which include higher-order amplitude squeezing and high-order photon
statistics \cite{nha}.

In this paper, we investigate in detail the entanglement generated via beam splitter
using two uncorrelated Gaussian input modes. In particular, we consider a
nonclassical single-mode state (squeezed state) and a thermal state at two
input modes, respectively. We note that Wolf {\it et al.} also considered a closely-related problem, i.e., 
they derived the conditions to generate bipartite Gaussian entanglement using passive transformations, focusing on the optimal scheme \cite{Wolf}. 
They proved that a 50:50 beam splitter is the optimal choice regardless of experimental parameters, and remarkably, 
that the optimal degree of entanglement depends on the smallest eigenvalues of the input covariance matrix.  
In other words, the degree of entanglement is solely determined by the degree of nonclassicality regardless of the purity of input state. 
In realistic situations, however, there always occurs an experimental error in designing beam-splitter, so a careful analysis of nonoptimal cases is further required.

In this respect, we first want to identify the whole experimental conditions to successfully generate
entangled Gaussian states. Second, a deeper issue is to establish
the link between single-mode nonclassicality and two-mode entanglement in general. 
A specific question we address is whether there exists a
monotonic relation between the degree of input nonclassicality and the
critical temperature (degree of noise) at which the output entanglement
disappears. For this purpose, we parametrize
an arbitrary single-mode Gaussian state in terms of the nonclassicality depth
$\tau$ \cite{Lee} and purity $u$, and study the behavior of entanglement as a function
of $\tau$, $u$, $\overline{n}$ ( thermal photon number at input), and the beam splitter transmittance. 
We demonstrate that the monotonic relation between the single-mode nonclassicality $\tau$ and 
the critical thermal noise $\overline{n}_c$ exists only at the optimal choice of 50:50 beam splitter 
and that $\overline{n}_c$ is generally a function of $u$ as well as $\tau$ for other choices of beam splitter.

This paper is organized as follows. In Sec.~II, a single-mode Gaussian state is briefly introduced with its covariance matrix 
in terms of nonclassical depth $\tau$ and purity $u$. 
In particular, the covariance matrix of the two-mode output state via a beam-splitter is obtained 
for the case that a nonclassical (squeezed) Gaussian state and a thermal state are used as two input modes. 
In Sec.~III, the degree of entanglement at the output is evaluated in terms of the logarithmic negativity and 
the experimental conditions to successfully generate entangled output are derived together with optimal setting. 
The critical thermal noise to destroy entanglement is analytically obtained and discussed in relation to the degree of single-mode nonclassicality. 
In Sec.~IV, our main results are summarized with concluding remarks.

\section{Two-mode states out of beam splitter}

Consider a lossless beam splitter whose input ports are fed by two single mode
fields with complex amplitude $\alpha_{1}$ and $\alpha_{2}$, respectively. The
complex amplitudes of the fields at the output ports are given by
\begin{equation}
\left(
\begin{array}
[c]{c}%
\beta_{1}\\
\beta_{2}%
\end{array}
\right)  =M_{B}\left(
\begin{array}
[c]{c}%
\alpha_{1}\\
\alpha_{2}%
\end{array}
\right)  ,
\label{eqn:trans}
\end{equation}
where $M_{B}$ is the beam splitter transformation matrix given as%
\begin{equation}
M_{B}=\left(
\begin{array}
[c]{cc}%
\cos\theta & \sin\theta e^{i\varphi}\\
-\sin\theta e^{-i\varphi} & \cos\theta
\end{array}
\right)  .
\label{eqn:transmat}
\end{equation}
The transmittance of the beam splitter is represented by $\cos^2\theta$ and
the phase difference between the reflected and the transmitted
fields by $\varphi$.

{\bf Nonclassical Gaussian state---}
Let the first input mode, $\alpha_1$, to the beam splitter be a single mode
Gaussian state defined by a characteristic function of the form%
\begin{equation}
\chi\left(  \mathbf{x}\right)  =\exp\left(  -\frac{1}{2}\mathbf{x}^{\dagger
}V_{1}\mathbf{x}\right), 
\label{eqn:GS}%
\end{equation}
where $\mathbf{x}^{\dagger}=\left(  \alpha_{1}^{\ast},\alpha_{1}\right)  $,
and $V_{1}$ is the covariance matrix%
\begin{equation}
V_{1}=\left(
\begin{array}
[c]{cc}%
a & b\\
b^{\ast} & a
\end{array}
\label{eqn:sing-cov}
\right)
\end{equation}
($a$: real, $b=\left\vert b\right\vert e^{i\phi}$: complex). In
Eq.~(\ref{eqn:GS}), we ignore the term linearly dependent on $\mathbf{x}$
which describes the displacement in phase space. This is because the local
displacement at each input emerges as another form of local displacements at
the output two modes so that it does not affect entanglement property at all.

A Gaussian state may be classical (coherent and thermal states) or
nonclassical (squeezed states). A number of measures have been proposed to quantify the degree of nonclassicality for a single-mode state \cite{Lee, Hillery, Knoll, Marian, Bures}. 
One of them, which will be used in this paper, is based on the
Glauber-Sudarshan $P$-function \cite{Glauber} defined as
\begin{align}
P(\zeta)  =\frac{1}{\pi^{2}}\int
d^{2}\alpha\chi(  \alpha)
e^{\frac{1}{2}|\alpha|^2-\alpha\zeta^{\ast}+\alpha^{\ast}\zeta},
\label{eqn:pfun}
\end{align}
where $\chi(\alpha)\equiv{\rm Tr}\{D(\alpha)\rho\}$ is the characteristic function. 
The $P$-function renders it possible to express the expectation values of normally ordered operator functions in close correspondence to the calculation of mean values in a
classical stochastic theory. A quantum state is said to have a classical
analog if its $P$-function has the properties of a classical probability
density. In general, however, the $P$-function may fail to be a probability distribution.
A quantum state is called nonclassical if its $P$-function is either singular or not positive-definite.

The integral in Eq.~(\ref{eqn:pfun}) may not be evaluated for a nonclassical state in general. 
However, a smooth and positive definite function that becomes acceptable as a
classical probability distribution is introduced by the convolution
transformation of the $P$-function \cite{Lee} as
\begin{equation}
R\left(  \tau,\eta\right)  =\frac{1}{\pi\tau}\int
d^{2}\zeta e^{-\frac{1}{\tau}\left\vert \zeta-\eta
\right\vert ^{2}} P(  \zeta).
\label{eqn:rfun}
\end{equation}
For a given $P$-function, there exists a certain value of $\tau_m$ such that the $R$-function becomes positive-definite for $\tau\ge\tau_m$.  
The threshold $\tau_m$ generally takes a value in $\left[0,1\right]$ and is regarded as a measure of
nonclassicality \cite{Lee}. 
In case of a Gaussian state with covariance matrix $V_1$, 
the condition for the positive definiteness of
$R\left(\tau,\eta\right)$ becomes 
\begin{equation}
V_{1}+\left(  \tau-\frac{1}{2}\right)  I>0,
\end{equation}
and $\tau_m$ thus takes a value in $\left[0,\frac{1}{2}\right]$. 
From now on, $\tau$ is used instead of $\tau_m$ to denote the nonclassical depth, and for the covariance matrix $V_{1}$ in Eq.~(\ref{eqn:sing-cov}), the degree of nonclassicality is given by
\begin{equation}
\tau=\max\left\{  0,-a+\left\vert b\right\vert +\frac{1}{2}\right\}.
\end{equation}

On the other hand, the degree of mixedness in a prepared quantum state
$\rho$ can be characterized by its purity $u={\rm tr}\left(  \rho
^{2}\right)$ ranging from $0$ (completely mixed state) to $1$ (pure
states). 
For a Gaussian state with covariance matrix $V_{1}$, the purity becomes \cite{Paris2003}%
\begin{equation}
u=\frac{1}{2\sqrt{\det V_{1}}}.
\end{equation}
In terms of the degree of nonclassicality $\tau$ and the purity $u$, one can thus
express the elements of the covariance matrix $V_{1}$ of a nonclassical Gaussian state as
\begin{align}
a  &  =\frac{1}{4u^2(1-2\tau)}+\frac{1}{4}(1-2\tau),\\
|b|  &  =\frac{1}{4u^2(1-2\tau)}-\frac{1}{4}(1-2\tau).
\end{align}
Although the parameter $b$ is complex, its phase $\phi$ does not play any role in the output entanglement [Eqs.~(\ref{eqn:neg}) and~(\ref{eqn:arg})]. Therefore, only two real parameters, $\tau$ and $u$, are sufficient to describe a general Gaussian state for our purpose.

{\bf Thermal state input---}
Let the second input mode, $\alpha_2$, to the beam splitter
be a thermal state defined as%
\begin{equation}
\rho_{\rm th}=%
{\displaystyle\sum\limits_{n}}
\frac{\overline{n}^{n}}{\left(  1+\overline{n}\right)  ^{n+1}}\left\vert
n\right\rangle \left\langle n\right\vert ,
\end{equation}
where $\overline{n}$ is the average photon number%
\begin{equation}
\overline{n}=\left[  \exp\left(  \frac{\hbar\nu}{k_{B}T}\right)  -1\right]
^{-1},
\end{equation}
where $k_{B}$ is the Boltzman constant and $T$ is the absolute temperature.
The thermal state is a classical Gaussian state with the covariance matrix
$V_{2}$ given by %
\begin{equation}
V_{2}=\left(
\begin{array}
[c]{cc}%
\overline{n}+\frac{1}{2} & 0\\
0 & \overline{n}+\frac{1}{2}%
\end{array}
\right)  ,
\end{equation}
with purity $u_{\rm th}=1/\left(2\overline{n}+1\right)  $.

{\bf Two-mode output---}
Having one-mode Gaussian state of nonclassicality $\tau$ and purity $u$ at one
port and thermal state at the other port of the lossless beam splitter, the
characteristic function of the two-mode input field can be written as %
$\chi\left(  \alpha_{1},\alpha_{2}\right)  =\exp\left(  -\frac{1}{2}%
\mathbf{w}^{\dagger}V_{\rm in}\mathbf{w}\right) $ ,
where $\mathbf{w}^{\dagger}\equiv\left(  \alpha_{1}^{\ast},\alpha_{1},\alpha
_{2}^{\ast},\alpha_{2}\right)  $ represents the complex amplitudes of
input modes and $V_{\rm in}\equiv V_1\oplus V_2$ the covariance matrix \cite{note}. %
On the other hand, the beam splitter action 
[Eqs.~(\ref{eqn:trans}) and (\ref{eqn:transmat})] 
yields the covariance matrix of the output characteristic function 
$\chi\left(  \beta_{1},\beta_{2}\right)  =\exp\left(  -\frac{1}{2}%
\mathbf{v}^{\dagger}V_{\rm out}\mathbf{v}\right) $ 
[$\mathbf{v}^{\dagger}\equiv\left(  \beta_{1}^{\ast},\beta_{1},\beta
_{2}^{\ast},\beta_{2}\right)$] 
as%
\begin{equation}
V_{\rm out}=\left( 
\begin{array}
[c]{cc}%
A & C\\
C^{\dag} & B
\end{array}
\right).
\label{eqn:cov}
\end{equation}
Here, the $2\times2$ matrices $A$, $B$, and $C$ are given by%
\[
A=\left(
\begin{array}
[c]{cc}%
a\cos^{2}\theta+\left(  \overline{n}+\frac{1}{2}\right)  \sin^{2}\theta &
b\cos^{2}\theta\\
b^{\ast}\cos^{2}\theta & a\cos^{2}\theta+\left(  \overline{n}+\frac{1}%
{2}\right)  \sin^{2}\theta
\end{array}
\right)  ,
\]%
\[
B=\left(
\begin{array}
[c]{cc}%
a\sin^{2}\theta+\left(  \overline{n}+\frac{1}{2}\right)  \cos^{2}\theta &
be^{-2i\varphi}\sin^{2}\theta\\
b^{\ast}e^{2i\varphi}\sin^{2}\theta & a\sin^{2}\theta+\left(  \overline
{n}+\frac{1}{2}\right)  \cos^{2}\theta
\end{array}
\right)  ,
\]%
\begin{equation}
C=\sin\theta\cos\theta\left(
\begin{array}
[c]{cc}%
\left(  a-\overline{n}-\frac{1}{2}\right)  e^{i\varphi} & be^{-i\varphi}\\
b^{\ast}e^{i\varphi} & \left(  a-\overline{n}-\frac{1}{2}\right)
e^{-i\varphi}%
\end{array}
\right).
\label{eqn:mat}
\end{equation}

\section{Quantitative measure of entanglement}
In this section, we study the degree of entanglement of the output two-mode states with the covariance matrix in Eq.~(\ref{eqn:cov}). 
A state described by a density operator $\rho$ is called separable if
it can be written as a convex sum of the product states, i.e., %
\begin{equation}
\rho=\sum_{i}p_{i}\rho_{A}^{i}\otimes\rho_{B}^{i},
\end{equation}
where $0\leqslant p_{i}\leqslant1$ and $\sum_{i}p_{i}=1$. Otherwise, it is
called entangled. A number of schemes have been proposed to verify quantum entanglement between two modes of the field \cite{Peres, Horo,
Simon, Werner, Duan}. In particular, it was shown that PPT criterion is sufficient and necessary for $1\times n$-modes bipartite Gaussian states. 
Instead of using these criteria, we consider
the quantitative measure of entanglement based on the logarithmic negativity
\cite{Vidal}.

\begin{figure}
\includegraphics*[width=2.1in,keepaspectratio=true]{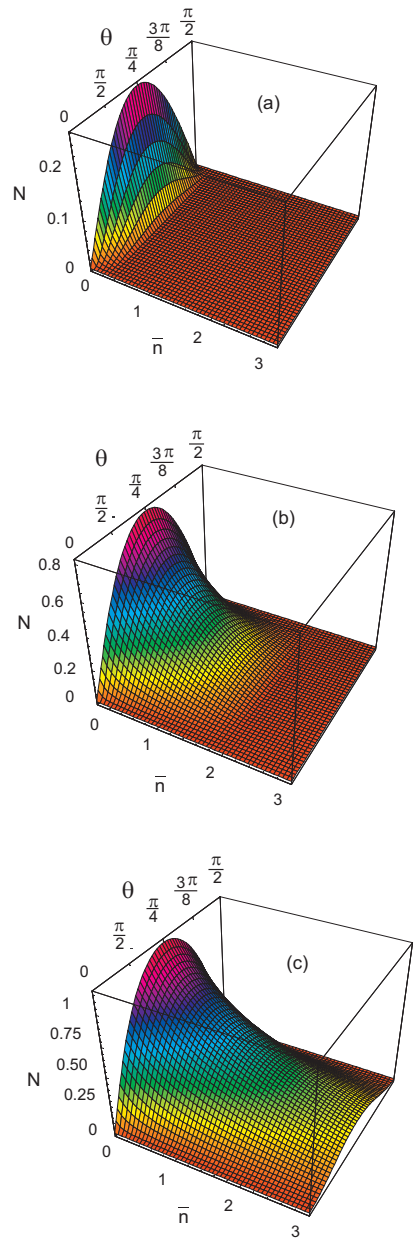}
\caption{Entanglement (logarithmic negativity) as a function of the thermal photon number $\overline{n}$ and the beam splitter angle $\theta$ 
for a pure-state Gaussian input ($u=1$) with nonclassical depth $\tau$. (a) $\tau=0.2$, (b) $\tau=0.4$, and (c) $\tau=0.45$.}
\label{fig:fig1}
\end{figure}

The logarithmic negativity is defined as $N\equiv\mathrm{log}_{2}||\rho^{\rm PT}||$, where $||A||\equiv{\rm tr}{\sqrt{A^\dag A}}$ denotes the trace norm and 
$\rho^{\rm PT}$ is the partially transposed density operator. 
For a general $n$-mode Gaussian state, 
the trace norm is determined by the 
eigenvalues of $-(V_r\sigma)^2$, the so-called symplectic eigenvalues of the real covariance matrix $V_r$. Here, the elements of the symplectic matrix $\sigma$ are given by the commutation relations, $[R_\alpha,R_\beta]=i\sigma_{\alpha\beta}$, where $R_\alpha$ ($\alpha=1,\cdots,2n$) denotes the canonical variables \cite{Simon1}. 
For the covariance matrix in Eq.~(\ref{eqn:cov}), the characteristic equation to evaluate the symplectic eigenvalues under partial transposition becomes \cite{Vidal}%
\begin{equation}
\xi^{4}-\left( {\rm Det} [A]+{\rm Det} [B]-2{\rm Det} [C]\right) \xi^{2}+{\rm Det} [V_{\rm out}]=0,
\end{equation}
where $A,B,$ and $C$ are the sub matrices of the matrix $V_{\rm out}$ in
Eq.~(\ref{eqn:cov}). 
Let two positive roots of this equation be $\xi_{\pm}$. 
The logarithmic negativity is then given by \cite{Vidal}
\begin{eqnarray}
N&=&\mathrm{max}\{0,-\mathrm{log}_{2}(2\xi_{-})\}+\mathrm{max}\{0,-\mathrm{log}_{2}(2\xi_{+})\}\nonumber\\ 
&=&\mathrm{max}\{0,-\mathrm{log}_{2}(2\xi_{-})\},
\label{eqn:neg-exp}
\end{eqnarray} 
because the condition ${\rm Det} [V_{\rm out}]={\rm Det} [V_{\rm in}]=\frac{(2\overline{n}+1)^2}{16u^2}>\frac{1}{16}$ always holds so that the larger root emerges as $2\xi_{+}\ge1$. 

Using Eqs.~(\ref{eqn:cov}) and (\ref{eqn:mat}), the negativity can be represented in terms of parameters $\tau$, $u$, $\overline{n}$, and $\theta$, 
which turns out to be
\begin{equation}
N={\rm max}\{0,-\frac{1}{2}\mathrm{log}_2\left(S-\sqrt{S^2-\frac{(2\overline{n}+1)^2}{u^2}}\right)\},
\label{eqn:neg}
\end{equation}
where 
\begin{eqnarray}
S\equiv \frac{1}{2}\left[\left(\overline{n}-\tau+1\right)S_+
-\left(\overline{n}+\tau\right)S_-\cos4\theta \right],\nonumber\\
\label{eqn:arg}
\end{eqnarray}
with
\begin{eqnarray}
S_{\pm}\equiv \frac{1}{u^2 (1-2\tau)}\pm (2\overline{n}+1).
\end{eqnarray}
Note that the negativity does not depend on the phase shift $\varphi$ at the beam splitter. 
In the following, we study in detail the degree of entanglement as a function of the experimental parameters.

\subsection {Optimal beam-splitter} 
In Eq.~(\ref{eqn:arg}), $S$ takes extremal values at $\theta=0$ and $\frac{\pi}{4}$. 
We obtain $S=S_0\equiv\frac{1}{2u^2}+\frac{(2\overline{n}+1)^2}{2}$ at $\theta=0$, and  
$S=S_{\frac{\pi}{4}}\equiv\frac{(2\overline{n}+1)}{2}\left(\frac{1}{u^2(1-2\tau))}+1-2\tau\right)$ at $\theta=\frac{\pi}{4}$. 
As $S_{0,\frac{\pi}{4}}$ are both positive, so is $S$ for the whole range of angles $\theta$. 
The logarithmic function in Eq.~(\ref{eqn:neg}) is a monotonically decreasing function of $S$ and the negativity therefore becomes maximal at the largest value of $S$.

For the case of $S_->0$, i.e., $\frac{1}{u^2 (1-2\tau)}>(2\overline{n}+1)$, in which the thermal photon number is relatively small, the maximum value of $S$ occurs at $\theta=\frac{\pi}{4}$ (50:50 beam splitter). 
On the other hand, for the case of $\frac{1}{u^2 (1-2\tau)}<(2\overline{n}+1)$, in which the thermal photon number is large, the maximum occurs at $\theta=0$, which essentially corresponds to no beam-splitter action and leads to no entanglement at all. Therefore, we conclude that the optimal choice of beam splitter is a 50:50 one regardless of all other parameters ( $\tau,u$, and $\overline{n}$.)

\subsection{Case of 50:50 beam splitter}

In this case, the negativity is reduced to 
\begin{equation}
N=\rm{max}\{0,-\mathrm{log}_2\sqrt{(2\overline{n}+1) (1-2\tau)}\}.
\label{eqn:optneg}
\end{equation}
The degree of entanglement is thus independent of purity, $u$, and depend only on the nonclassicality, $\tau$. 
From Eq.~(\ref{eqn:optneg}), the critical thermal noise for the vanishing negativity, $N=0$, is obtained as 
\begin{eqnarray}
\overline{n}_c=\frac{\tau}{1-2\tau}.
\label{eqn:critic}
\end{eqnarray}
If $\overline{n}\ge\overline{n}_c$, the entanglement at the output disappears. 
Note that the critical value $\overline{n}_c$ is a "monotonic" function of nonclassicality $\tau$ regardless of purity $u$. 
Therefore, the resistance to noise, $\overline{n}_c$, in generating entangled output can be understood as equivalent to a measure of single-mode nonclassicality, $\tau$. 
At the maximal squeezing, $\tau\rightarrow\frac{1}{2}$, the critical value approaches $\overline{n}_c\rightarrow\infty$, i.e., entanglement persists at any level of noise.

\begin{figure}
\includegraphics*[width=2.4in,keepaspectratio=true]{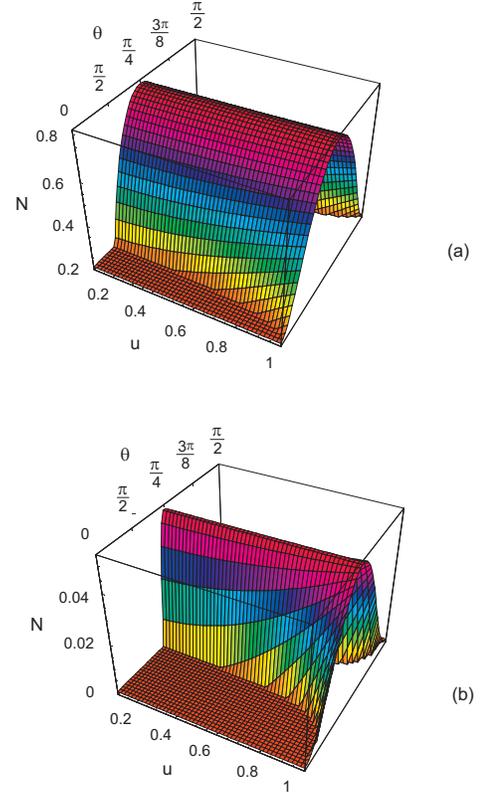}
\caption{Entanglement (logarithmic negativity) as a function of the purity $u$ and the beam splitter angle $\theta$ 
at a fixed level of thermal noise $\overline{n}$ for the nonclassical depth $\tau=0.45$. (a) $\overline{n}=1$ and (b) $\overline{n}=4$.}
\label{fig:fig2}
\end{figure}

\subsection{Case of general beam splitter}

For a general beam-splitter angle $\theta$, the critical value $\overline{n}_c$ is obtained by requiring $2S=1+\frac{(2\overline{n}+1)^2}{u^2}$ in Eq.~(25). 
Unlike the case of 50:50 beam splitter, the critical noise $\overline{n}_c$ and the logarithmic negativity $N$ depend on the initial purity $u$ as well as the nonclassicality $\tau$ of the input state. 
This implies that the interpretation of the critical noise as a measure of nonclassicality is not valid for a general beam-splitter setting. 
For example, at the choice of $\theta=\frac{\pi}{12}$, the critical value becomes $\overline{n}_c=0.75$ for $\tau=0.3$ and $u=1$ (pure-state), and 
$\overline{n}_c\approx0.36$ for $\tau=0.4$ and $u=0.2$ (mixed-state). 
In this example, the higher nonclassical depth leads to the lower critical thermal noise. 

{\bf Case of near-optimal BS:} Let us denote the beam splitter angle by $\theta=\frac{\pi+\delta}{4}$, where $\delta$ is a small error. 
The fractional deviation from the optimal transmittance 1/2 then corresponds to $e\equiv\frac{\delta}{2}$, and the critical thermal noise is found to be 
\begin{eqnarray}
\overline{n}_c\approx\frac{\tau}{1-2\tau}\left(1-2e^2\frac{(1-\tau)(1-u^2)}{1-u^2(1-2\tau)^2}\right),
\label{eqn:nearoptimal}
\end{eqnarray}
which shows the dependence on the purity as well as the nonclassicality of the input state. 
In case of very small squeezing, $\tau\ll 1$, the dependence on the purity is negligible as $\overline{n}_c\approx \tau\left(1-2e^2\right)$. 
On the other hand, close to maximal squeezing, $\tau\rightarrow\frac{1}{2}$, we obtain $\overline{n}_c\approx\frac{1}{2-4\tau}\left(1-e^2(1-u^2)\right)$. 

In the following, we consider in more detail the case of general BS angles.

{\bf Pure-state input:} Let us consider the case that the nonclassical resource at the input is pure, i.e., $u=1$. 
we plot the negativity as a function of beam-splitter angle $\theta$ and the thermal photon number $\overline{n}$ 
for a fixed value of $\tau$ (degree of nonclassicality) in Fig.~1.
Obviously, the entanglement becomes optimal for the choice of $\theta=\frac{\pi}{4}$ and decreases as the angle deviates from $\frac{\pi}{4}$. 
In this case, one can obtain the analytic expression of critical thermal noise as $\overline{n}_c=\frac{\tau}{1-2\tau}$, which is remarkably independent of angle $\theta$. 
Therefore, although the degree of entanglement varies with the beam-splitter parameter $\theta$, the entanglement disappears at the same level of noise $\overline{n}_c=\frac{\tau}{1-2\tau}$ regardless of $\theta$.

{\bf Mixed-state input:}
In general, the degree of entanglement grows with increasing nonclassical depth $\tau$ and purity $u$. 
In Fig.~2, the logarithmic negativity is plotted as a function of the purity $u$ and the beam-splitter angle $\theta$ at a fixed level of thermal noise. 
We see that a successful generation of entangled output occurs in a broader range of angles $\theta$ with increasing purity $u$. 
In Fig.~3, the critical noise $\overline{n}_c$ is plotted as a function of purity $u$ and the beam-splitter angle $\theta$ for a fixed nonclassical depth $\tau=0.4$. 
As the purity $u$ increases, the distribution of critical value $\overline{n}_c$ becomes broader with respect to the beam-splitter angle, and it eventually becomes flat at $u=1$, as argued in the previous paragraph. 

\begin{figure}
\includegraphics*[width=3.2in,keepaspectratio=true]{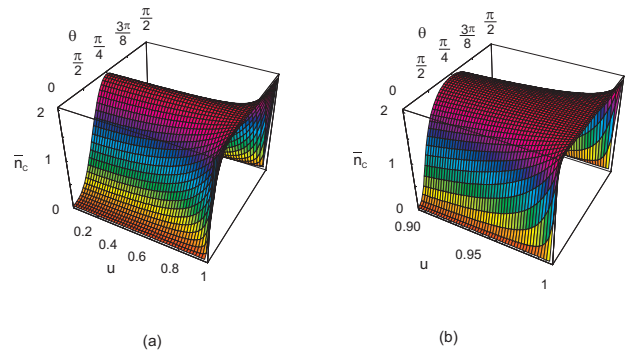}
\caption{The critical thermal noise $\overline{n}_c$ as a function of the purity $u$ and the beam splitter angle $\theta$  
for a fixed nonclassical depth $\tau=0.4$. The right plot (b) shows a magnified view over a narrow range of $u$ close to 1.}
\label{fig:fig3}
\end{figure}

\section{Conclusions and Remarks}
In summary, we have investigated in detail the generation of entangled Gaussian states via a beam splitter using a single-mode squeezed state and a thermal state as two inputs. We have identified the condition to successfully produce an entangled state at the output, and evaluated the degree of entanglement as a function of experimental parameters, i.e., the nonclassical depth $\tau$ and the purity $u$ of nonclassical source, the thermal photon number $\overline{n}$, and the beam-splitter angle $\theta$ (transmittance). 

We have established the connection between the critical thermal noise $\overline{n}_c$ and the nonclassical depth $\tau$, and 
showed that the connection gains a strong interpretation only at the optimal choice of 50:50 beam splitter. 
In other cases ($\theta\ne\frac{\pi}{4}$), the critical noise is a function of the purity $u$ as well as the nonclassical depth $\tau$ 
so that a higher nonclassicality does not necessarily lead to a more robust resistance to thermal noise. 


It is noteworthy that the critical noise in Eq.~(\ref{eqn:critic}) to destroy output entanglement coincides with the amount of thermal noise that can be introduced to the input Gaussian state to destroy its single-mode nonclassicality (squeezing) in a specialized setting: Suppose one starts with a vacuum state $|0\rangle$ to produce a mixed squeezed state $\rho$ that has the covariance in Eq.~(\ref{eqn:sing-cov}) and the nonclassicality $\tau$. In general, $\rho$ can be expressed in the Kraus-sum representation as $\rho=\sum_iM_i|0\rangle\langle0|M_i^\dag$, where $\sum_iM_iM_i^\dag=I$. 
Now, if the initial vacuum state is replaced by a thermal state with the photon number $\overline{n}_{\rm th}$ as $\rho'=\sum_iM_i\rho_{\rm th}M_i^\dag$, 
it is easy to show that the single-mode state $\rho'$ becomes classical at $\overline{n}_{\rm th}=\frac{\tau}{1-2\tau}$, which is none other than the critical noise in Eq.~(\ref{eqn:critic}). Therefore, the two contextually different noises coincide quantitatively in the Gaussian regime. Of course, the result $\overline{n}_{\rm th}=\frac{\tau}{1-2\tau}$ is valid only for $\tau<\frac{1}{2}$, and thus cannot be readily extended to non-Gaussian regime, e.g. Fock-states ($\tau=1$). 
Nevertheless, it seems plausible to have such a relation even for non-Gaussian states in a different form. 

We also note that C. T. Lee attempted to connect the nonclassical depth $\tau$ to the thermal photon number required to destroy all nonclassical aspects of the state \cite{Lee}. However, the link by C. T. Lee is rather {\it formal}, and precisely speaking, the parameter $\tau$ in Eq.~(\ref{eqn:rfun}) is a Gaussian noise \cite{Hall} not necessarily arising from a thermal state: When an initial state $\rho$ is exposed to a Gaussian noise as
\begin{eqnarray}
\rho'=\int d^2\beta P(\beta) D(\beta) \rho D^\dag(\beta),
\end{eqnarray}
where the state $\rho$ is displaced in phase space by the amount of $\beta$ with the Gaussian weighting  
$P(\beta)\equiv\frac{1}{\pi\sigma} e^{-\frac{|\beta|^2}{\sigma}}$, the $P$-function of the output state $\rho'$ is none other than the convolution in Eq.~(\ref{eqn:rfun}), with the identification $\sigma=\tau$. 
(Note that this Gaussian noise is different from the noise process mentioned in the preceding paragraph.) 
On the other hand, if one mixes the initial state with a thermal state with at 50:50 beam-splitter, then output state possesses the $P$-function as 
$P(\eta)=2R\left(\sqrt{2}\eta\right)$, that is, a scaled distribution of the $R$ function of Eq.~(\ref{eqn:rfun}). 
Not to mention that the scaled function represents a different density operator, 
it is also known that a scaling transform in phase space does not generally map a physical state to another physical one \cite{nha2,nha3}. 
In contrast, our connection of $\tau$ to critical thermal noise has a clear physical meaning in a realistic experimental scheme. 
Our scheme is, however, restricted to the class of Gaussian states and it is thus desirable to extend the current issues to non-Gaussian states in future. 


\section{Acknowledgment}

Authors thank COMSTECH, Pakistan for their support. HN and MSZ are supported by an NPRP grant 1-7-7-6 
from Qatar National Research Fund. Authors greatly acknowledge M. M. Wolf and J. Eisert for alerting them with the paper \cite{Wolf} with stimulating discussions.  
*email: hyunchul.nha@qatar.tamu.edu

\appendix

\begin{center}

\end{center}


\begin{thebibliography}{99}
\bibitem{note} In this paper, complex-valued covariance matrices, $V_c$, are constructed with respect to field amplitudes $\alpha$ and $\alpha^*$, 
whereas in other literatures real-valued ones, $V_r$, are usually constructed with respect to canonical variables $x$ and $p$, e.g. \cite{Simon}. 
Using the relation $\alpha=\frac{1}{\sqrt{2}}(x+ip)$, however, one can easily translate one into the other as $V_r=C^\dag V_cC$ where $C\equiv C_1\oplus C_1$ is a unitary matrix with 
\begin{equation}
C_{1}=\frac{1}{\sqrt{2}}\left(
\begin{array}
[c]{cc}%
1 & i\\
1 & -i
\end{array}
\right).  \nonumber
\end{equation}
\bibitem{nha3} Furthermore, the convolution of two $P$ functions, $P(\eta)=\int d^2\zeta P_1(\zeta) P_2(\eta-\zeta)$ as in Eq.~(\ref{eqn:rfun}), 
which reads as $C(\xi)=e^{\frac{1}{2}|\xi|^2} C_1(\xi)C_2(\xi)$ at the level of characteristic function, does not necessarily represent a physical state \cite{Hall}. 
For example, take one-photon Fock states, $\rho_{1,2}=|1\rangle\langle1|$. 
It is straightforward, using the method described in \cite{nha2}, to obtain the resulting quasi-density operator as $\rho=|0\rangle\langle0|-2|1\rangle\langle1|+2|2\rangle\langle2|$, which is unphysical.
\end{thebibliography}
\end{document}